# High Intensity for Accelerator Driven Systems (ADS)

*U. Dorda*
SCK CEN, Mol, Belgium

**Abstract**
Following the motivation for an Accelerator Driven System (ADS), the requirements on the accelerator are derived. Using the MYRRHA project as example, the beam optics/dynamics design and operational concept of such an accelerator are discussed and the main technology choices and challenges are presented.

## 1 Introduction

### 1.1 Motivation

Given the urgent need to limit climate change and meet rising global energy demand, nuclear reactors continue to play a key role in electricity generation across many countries. By the end of 2023, 417 reactors were in operation across 31 countries, with 58 more under construction. Nuclear power accounted for approximately 10% of global electricity production, rising to 25% within the European Union and nearly 50% in Belgium. At COP28, a global declaration was signed to triple nuclear energy capacity by 2050, underscoring its critical role in achieving net-zero emissions.

An Accelerator-Driven System (ADS) is a technology designed to address several major drawbacks of conventional fission reactors.

### 1.2 Thermal-neutron reactors

Thermal-neutron reactors are currently the predominant type of nuclear reactor in operation. However, they face several key challenges:

– Accumulation of long-lived nuclear waste

– Limited availability of nuclear fuel: Current estimates suggest that conventional U-235 resources used in today's reactors could be depleted by the end of this century

– Inherent risks of operating a critical reactor

– Proliferation concerns linked to weapons-grade material.

While the initial natural nuclear fuel is already radioactive, its radiotoxicity is rather low. Radiotoxicity is a measure of the potential harm caused by radioactive substances once they are taken into the body. It is distinct from chemical toxicity and specifically refers to the damage caused by the radiation emitted as these substances decay.

For example:

– Tritium has low radiotoxicity. It emits low-energy beta particles (less than 18 keV) and is typically eliminated from the body before it has a chance to decay and cause harm.

- Plutonium, by contrast, has high radiotoxicity. It is an alpha emitter and tends to accumulate in the bones and liver, where it can remain for a long time, posing a significant long-term health risk.

Figure 1 illustrates the typical nuclear reaction occurring within a thermal-neutron reactor. In this setup, water serves a dual purpose: it extracts heat generated by fission and acts as a moderator to slow down neutrons. By reducing the kinetic energy of neutrons to thermal levels (approximately 0.025 eV), the probability of inducing fission increases significantly - reflected in a much higher cross-section of about 1000 barns, compared to just 1 barn for fast neutrons (approximately 1 - 2 MeV).

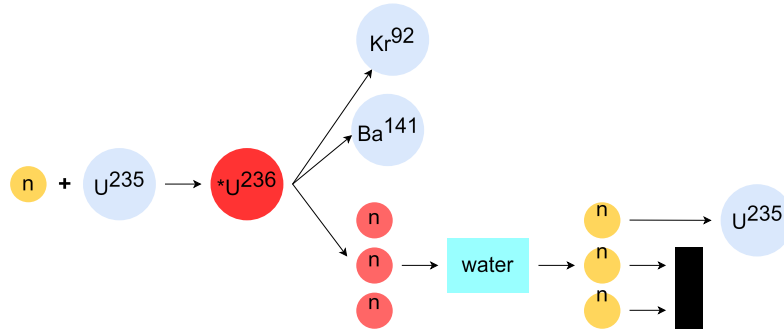

**Fig. 1:** yellow = thermal neutron, red = fast neutron, black = Neutron absorbers: boric acid added to water or/and mechanical control rods.

If this were the sole reaction taking place within the reactor, two notable consequences would arise:
- Reactor control would be ineffective if all neutrons were prompt, as the system's response in the μs time scale would be too short for control mechanisms - such as control rods - to act effectively. This would severely hinder the ability to stabilize or regulate the reactor's power output. In reality, however, a small fraction of neutrons, known as delayed neutrons, are emitted seconds after fission due to the decay of certain fission products (e.g., Ba-141). In a typical pressurized water reactor (PWR), approximately 0.65% of the neutrons are delayed, which plays a critical role in enabling safe and manageable reactor control.
- Radioactive waste would pose only a minor concern if limited to the primary fission products, such as krypton (Kr) and barium (Ba), whose radiotoxicity declines to levels comparable to natural uranium ore within about 100 years. However, in reality, several other decay pathways (see Fig. 2) produce long-lived, highly radiotoxic elements - most notably the minor actinides. These include isotopes with half-lives of hundreds of thousands of years. For context, one ton of spent nuclear fuel, which can supply electricity to approximately 100,000 households for 4.5 years, contains only a few kilograms of these long-lived actinides. Despite their small quantity, they represent a significant long-term challenge in radioactive waste management.

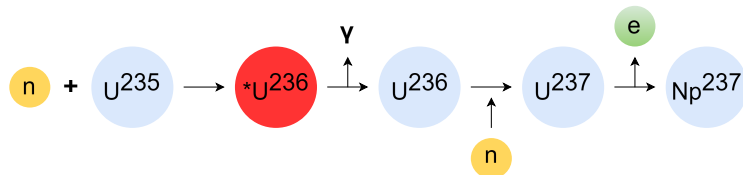

**Fig. 2:** Example of a decay channel creating the minor Actinide Np.

## 1.3 Fast reactors

Fast reactors are not widely deployed, primarily due to their increased complexity and associated costs. They do not require a moderator as they operate using unmoderated (fast) neutrons. Although their coolants pose challenges - such as flammability in the case of sodium, and corrosion in lead-bismuth eutectic - they benefit from operating at atmospheric pressure and have very high boiling points, enhancing thermal efficiency and safety margins. One of the major advantages of fast reactors is their ability to efficiently fission the much more abundant uranium-238, effectively removing fuel supply limitations. They can also transmute and fission minor actinides, helping to reduce long-lived nuclear waste; however, safety constraints significantly limit the allowable concentration of these isotopes in the fuel due to e.g. the lower proportion of delayed neutrons.

## 1.4 ADS concept

An ADS concept is a fast reactor with a subcritical reactor which cannot sustain a stable chain reaction on its own. A high-power particle accelerator is coupled into the core to the "missing" fast neutrons (20 - 30 neutrons per proton). The advantages are:

- Inherent Safety: Since the core is subcritical, the reaction automatically stops if the accelerator is turned off.
- Waste Transmutation: An ADS can efficiently burn long-lived radioactive waste, including minor actinides.
- Fuel Flexibility: Capable of using thorium or spent nuclear fuel, making better use of available resources.

On the other hand, there is the challenge to produce the reliable, high-power accelerator that can operate continuously and economically.

The proposed reactor designs use Lead-Bismuth Eutectic (LBE, eutectic alloy of 44.5 % lead and 55.5 % bismuth) for the following reasons:

- Neutronic Transparency: The high atomic mass makes LBE nearly "transparent" to neutrons, minimizing parasitic absorption and maximizing neutron economy in the reactor core.
- Thermal Properties: Low melting point (~125°C) and high boiling point (~1670°C) allow operation over a wide temperature range, simplifying thermal management and safety.
- Spallation Target: LBE can directly serve as a spallation target, generating neutrons when bombarded by high-energy protons.

## 2 Accelerator requirements

The neutron creation yield per MeV of incoming proton energy (see Fig. 3) increases steeply for proton beam energies ($E_{beam}$) below 500 MeV and then flattens. For the yields reached at >500 MeV, a proton beam with MW-class average proton beam power ($P_{avg}$) is required to generate the required neutron-flux to keep the power level of the sub-critical reactor core stable. While a $P_{avg}$ of 2 to 3 MW suffices to demonstrate the transmutation concept and the feasibility of an ADS (e.g. 2.4 MW at 600 MeV for the MYRRHA reactor with a 70 MW thermal output power), future industrial ADS for power generation to the grid will most likely require $P_{avg}$ > 10 MW (e.g. 30 MW in case of the JAEA-ADS design study for an 800 MWth energy resulting in 170MW delivered to the grid and 100 MW to run the accelerator).

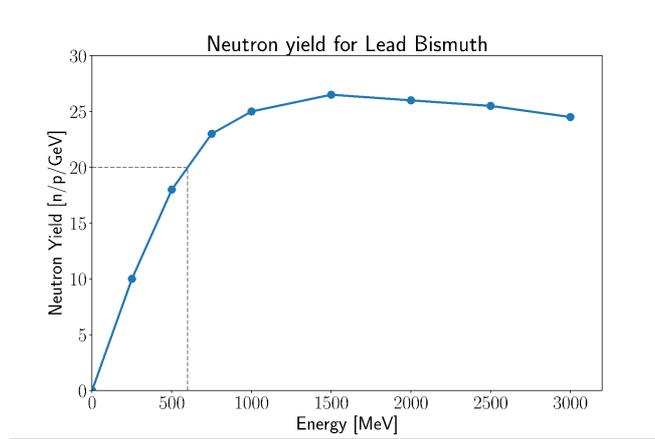

**Fig. 3:** Neutron yield per GeV of incoming proton energy for Lead Bismuth, data from:A Zaetta, "Transmutation is technically feasible", 2022. The dotted lines indicate the MYRRHA working point.

To minimize space charge effects, enable the use of solid-state amplifiers as RF sources, and reduce thermal stress on the target window, continuous wave (CW) beam operation is selected. Additionally, incorporating the option to adjust the duty factor allows for tuning of the power delivered to the reactor, providing greater operational flexibility and control.

While a few high-power accelerators currently in operation worldwide are capable of meeting the necessary beam parameters listed in Table 1, achieving the required levels of availability and reliability remains a key focus of ongoing research in order to limit the thermal stress in the ADS and to provide stable energy production in case of an industrial ADS.

**Table 1:** Main accelerator specifications

| Parameter | Value |
| --- | --- |
| Particle type | Protons |
| Final Energy (MeV) | > 500 |
| Beam power | MW level |
| Beam pattern | Up to DC |
| Availability | Quasi continuous |
| beam trips longer than a few seconds per 90 days | < 10 |

## 3 Accelerator options

There are two accelerator types capable to meet these requirements:

- Cyclotrons: Since its initial commissioning in 1974, the 600 MeV separated-sector cyclotron at PSI has undergone continuous upgrades and now routinely delivers beam currents of up to 2.2 mA, demonstrating the power capabilities of this accelerator type. Several ADS design studies have utilized cyclotron technology [3], and various commercial companies are also exploring related applications [4].

- Superconducting linacs: The SNS [8] accelerator complex has been operational since 2006, delivering up to 1.4 MW of beam power with an upgrade to 2.8 MW currently being underway. While also bringing challenges, beam currents of multiple tens of mA have been routinely demonstrated and energy scaling – while financially costly – is straight forward.

In addition to numerous superconducting linac-based design studies - such as the JAEA-ADS - two major projects are currently in implementation:

- CiADS using 6 different types of cryomodules
- MYRRHA: the 1st stage (100 MeV) is currently being implemented. This accelerator design will be used in the following as reference case.

The following covers only the superconducting linac based option.

## 4  Accelerator overview

The linac can typically be split into three distinct sections (see Fig. 4):

- A normal conducting injector
- A superconducting section
- A beam delivery line to the reactor.

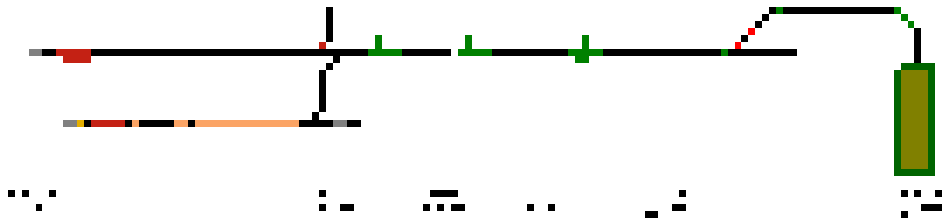

**Fig. 4:** Not-to-scale layout of the MYRRHA accelerator. Two injectors are foreseen to allow continued beam delivery in case of failure, yellow/blue = normal/super-conducting RF cavities.

The transition from normal-conducting to superconducting is dictated by an overall optimization: While normal-conducting RF structures have high resistive power loss, superconducting cavities are significantly more complex and sensitive to operate and require a costly and energy-hungry cryo-plant. Compared to pulsed high power linacs, the required CW RF operation makes the superconducting option favourable already at relatively low energies: at 2 MeV in case of the CiADS and 17 MeV in case MYRRHA ADS (compared to 100 MeV in the pulsed ESS accelerator).

## 5  RF cavity & frequency choices

The RF frequencies are chosen based on requirements by beam dynamics (e.g. transition time factor, required apertures) and the availability of efficient RF sources and structures. For MYRRHA, 4 different sections (and related RF structure designs) have been identified and are detailed below.

### 5.1.1  176 MHz normal conducting injector up to 17MeV

#### 5.1.1.1  Radio-Frequency Quadrupole (RFQ)

There are two main design types of RFQs:

- 4-vane RFQ designs are the only viable option at frequencies of several hundred MHz. Their RF properties depend not only on the vane geometry but also significantly on the cavity wall shape, making them highly sensitive to mechanical tolerances.

- 4 rod designs become attractive at lower frequencies due to their increased tolerance to mechanical imperfections. Since the shunt impedance of RFQs scales approximately with $1/f^{1.5}$, the 4-rod configuration is particularly suitable at 176.1 MHz - the frequency chosen for the MYRRHA injector. For practical manufacturing reasons, the total structure length was limited to 4 meters. While it is designed to handle up to 160 kW of RF power, 120 kW are used in routing operation to accelerate the beam to up to 1.5 MeV.

*5.1.1.2 CH-cavities*

Acceleration from 1.5 to 17 MeV is achieved using 15 accelerating and 4 re-bunching normal-conducting "Cross-bar H-mode" (CH) cavities. CH structures (see Fig. 5) are based on the drift tube linac concept, with the TE211 quadrupole mode resonating in π-mode inside a cylindrical cavity. Multiple stems - at least four per cavity - support the drift tubes and are arranged with alternating 90° rotations.

At these relatively low energies, proton velocity still varies significantly along the linac, requiring each cavity to have a unique design - varying in diameter, stem length, and number of drift tubes.

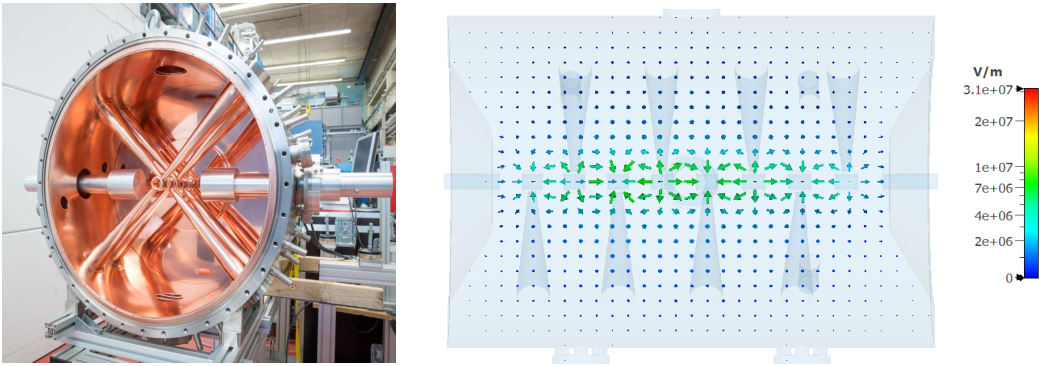

**Fig. 5:** CH cavity; Left: Photo of an open CH-cavity as designed by IAP Frankfurt, Right: Electric Field distribution inside a cavity.

As the CH-design allows for relatively small apertures, CH cavities achieve a higher shunt impedance compared to traditional Alvarez or Interdigital H-mode (IH) linacs. While a small aperture may be an issue for beam loss, this makes them more energy-efficient and compact. However, despite these advantages, only 70 kW of the 700 kW of RF power is transferred to the beam. This is illustrated in Fig. 6, where the beam power added by each cavity is plotted together with the power which needs to be supplied by the power supply to each CH cavity. At the position of the re-bunching cavities, no beam power is added. The synchronous phases and the cavity field gradient settings in all CH cavities are shown in the top part of Fig. 14.

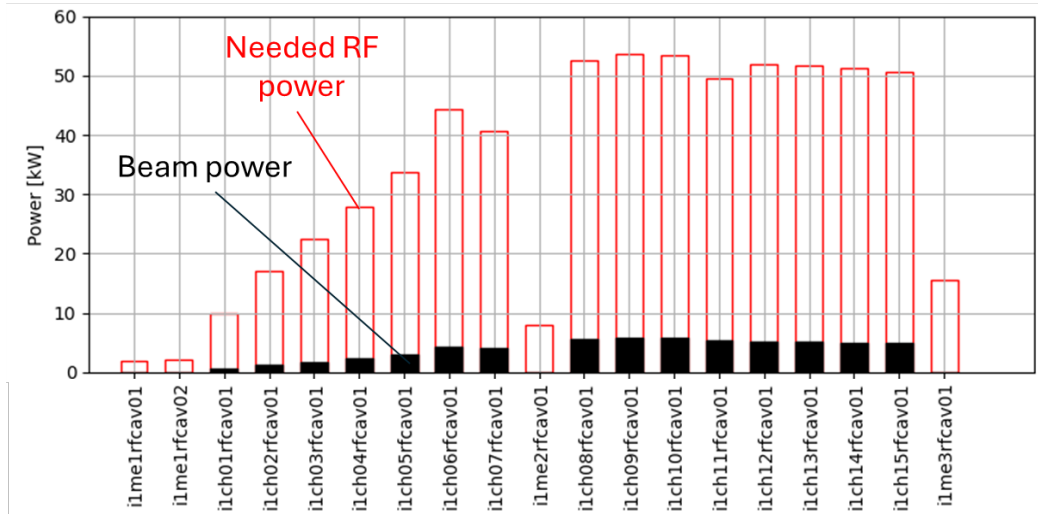

**Fig. 6:** The beam power and the needed RF power to the CH cavities. The cumulative sum of all RF power is approximately 700 kW, whereas the cumulative beam power is only 70 kW.

Figure 7 plots the transit time factor (T) of the CH-cavities against input beam energy. Although they achieve high efficiency at their design working point (T > 0.8), performance declines sharply at nearby energies due to the rapid velocity change at non-relativistic speeds and the multi-cell structure. As a result, the failure of a single cavity cannot be mitigated by retuning adjacent cavities. Keeping in mind that reliability is a key in order to limit the thermal stress in the ADS and to provide stable energy production in case of an industrial ADS, this is then one of the main reasons for having a second injector, as shown in Fig. 4.

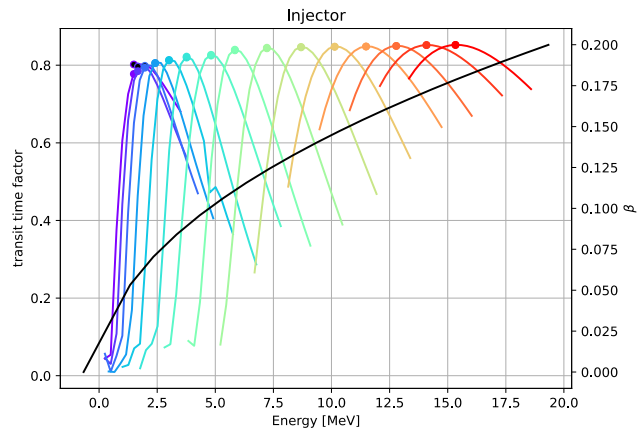

**Fig. 7:** Transit time factor for the different CH cavities as a function of the incoming beam energy. The nominal working points are indicated by the dots.

### 5.1.2    352 MHz and 704 MHz superconducting RF main linac

Given the challenges in designing and fabricating superconducting cavities and their cryomodules, minimizing the number of distinct cavity types is essential. For MYRRHA, three superconducting cavity designs are planned:

–    For the low-beta region single spoke cavities and for high-beta elliptical cavities.

- For the medium-beta range, two options are under evaluation for the medium-beta section (double-spoke or elliptical cavities). Elliptical cavities are preferred for the high beta section.

Single-Spoke Cavities (see Fig. 9) are half-wave cavities, featuring a transmission-line segment (the "spoke") mounted perpendicular to the cylindrical outer housing. This design enables operation at relatively low frequencies while keeping the cavity size practical. With only a few cells per cavity, the transit time factor remains sufficiently high over a broad velocity range (see Fig. 8 for the MYRRHA single-spoke cavity example). Multi-Spoke Cavities are composed of several spokes (typically up to three), each rotated 90° relative to the others. Although the accelerating field of spoke cavities may appear modest, the surface field is comparable to that of the best elliptical cavities. However, spoke cavities are more complex to manufacture and clean than elliptical ones.

**Table 2:** Summary of cavity types and settings for the MYRRHA linac.

| Section # | Low beta | Medium beta Option 1 | Medium beta Option 2 | High beta |
|---|---|---|---|---|
| $E_{kin,in}$ [MeV] | 16.6 | 80.8 | 80.8 | 172.3 |
| $E_{kin,out}$ [MeV] | 80.8[1] | 172.3 | 172.3 | 601.6 |
| Technology | Single Spoke | Double spoke | Elliptical | Elliptical |
| Frequency [MHz] | 352.2 | 352.2 | 704.4 | 704.4 |
| optimal β | 0.351 | 0.495 | 0.510 | 0.705 |
| Nb of cells / cavity | 2 | 3 | 5 | 5 |
| Nb of cavities / cryomodule | 2 | 2 | 2 | 4 |
| Nb of cavities | 48 | 28 | 34 | 72 |
| Nominal Eacc (MV/m) | 7.0 | 6.8 | 8.2 | 11.0 |
| Max Eacc (MV/m) (fault recovery) | 9.1 | 9.0 | 10.7 | 14.3 |
| Synch. phase (deg) | -45 to -15 | -35 to -15 | -36 to -15 | -36 to -15 |

---

[1] Due to a staged implementation approach of MYRRHA requiring 100 MeV beam to intermediate users, the low beta linac energy range is actually extended to 100 MeV using a total of 60 single spoke cavities.

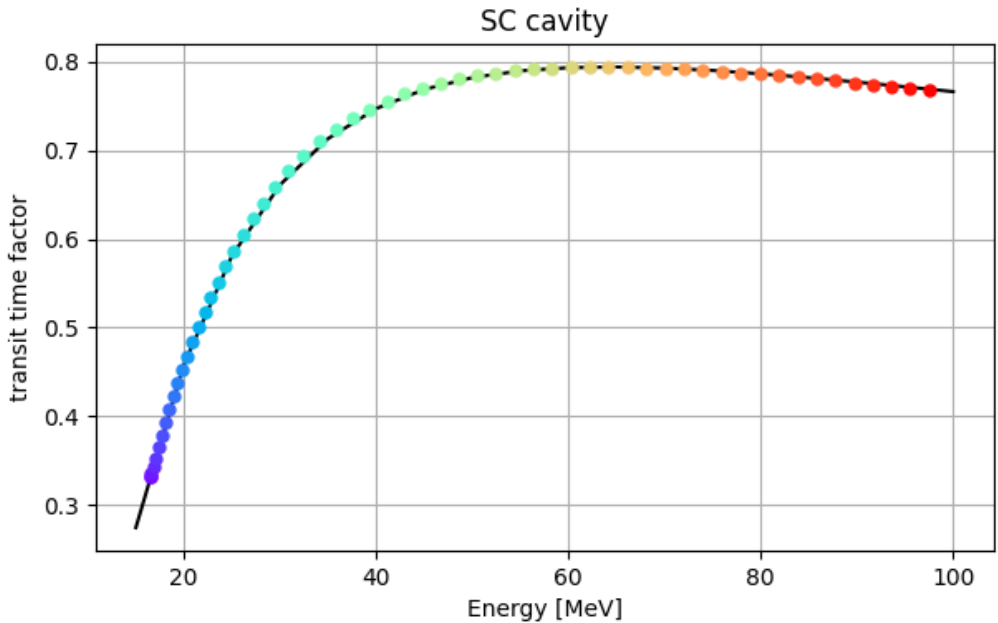

**Fig. 8:** Transit time factor for the super conducting single spoke cavities as a function of beam energy. The colour dots refer to the input energies of the 60 individual single spoke cavities of the LB linac in MYRRHA phase 1.

Superconducting cavities are characterized by an exceptionally high-quality factor (Q), which has

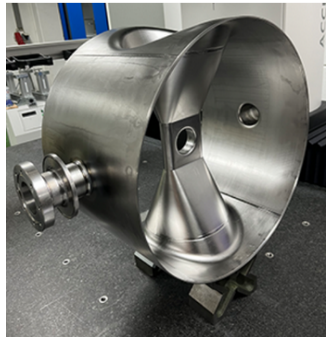

several important implications:

- A high Q allows for a larger aperture, improving beam optics and significantly reducing beam loss.
- Reflected Power: When RF is present but beam is not (e.g., during duty factor adaptation or RF-filling), nearly all RF power is reflected. This necessitates circulators in RF transmission lines to protect RF sources.
- Filling Time: Filling cavities with RF takes longer compared to normal conducting cavities.
- Rapid Detuning: If a cavity or related system (RF generator, LLRF) fails, it must be mechanically detuned using the cold tuning system. For an ADS, this must occur within the reconfiguration time (e.g., <1s). Failure to detune can lead to:
    - Cavity Damage: Excessive power deposition in the cavity.
    - Beam Energy Loss: Power induced in the cavity is drawn from the bunch.
    - Bunch Quality Degradation: Induced voltage negatively impacts subsequent bunches.

- The Low-Level Radio Frequency (LLRF) system must precisely control the cavity's resonance frequency to maintain performance and stability.

In RF-pulsed cavities, quench detection typically occurs during the RF ramp-down, allowing an inhibit signal to block the next pulse if necessary. However, continuous-wave (CW) RF operation demands real-time quench detection while the RF is actively applied.

# 6 Accelerator layout and beam physics

## 6.1 Injector

Figure 9 shows the layout of the first part (up to the 7$^{th}$ accelerating CH cavity out of 15 accelerating cavities in total) of the MYRRHA injector.

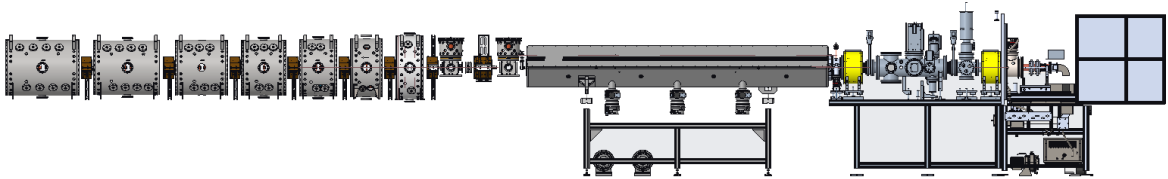

**Fig. 9:** Layout of the MYRRHA injector up to the 7th CH cavity, beam travelling from right to left. The HV-cage of the ion source is shown retracted here to allow a view of the ion source.

Commercial ECR ion sources reliably deliver the required 30 keV beam at specified current levels. The Low Energy Beam Transport (LEBT) line serves several critical functions:

- Particle Filtering: Removes undesired particle species emitted from the ion source.
- Beam Focusing: Uses two solenoids to match the beam into the RFQ acceptance (see Fig. 10).
- Beam Diagnostics: Measures beam current (Faraday cup or ACCT) and emittance (Allison scanner).
- Beam Structuring: Employs an electrostatic chopper (rise time <1 µs) to shape the macro-beam structure.

In the LEBT, Space Charge Compensation can occur: Residual gas in the LEBT is ionized by the proton beam and while protons are repelled, the resulting electron cloud mitigates space charge effects. For MYRRHA's beam current, the need for intentional gas injection (e.g., Argon) is still under evaluation.

When the chopper voltage is ramped down to permit beam passage, the space charge compensation builds up over tens of microseconds. During this transient period, the transverse focusing—and consequently the beam matching—is temporarily altered.

The LEBT is followed by an RFQ, which focuses and bunches the beam at 176.1 MHz and accelerates it to 1.5 MeV.

In the subsequent Medium Energy Beam Transfer Section 1 (MEBT1), two quarter-wave resonators and quadrupole doublets ensure longitudinal and transverse beam matching into the following CH section.

The CH section comprises 15 accelerating and 3 re-bunching CH-cavities, each optimized for the beam's energy. Between cavities, quadrupole doublets transport the beam, whereas orbit correctors and BPMs are used for to maintain beam on axis. The 6σ beam sizes in the injector part of the MYRRHA linac are shown in Fig. 11.

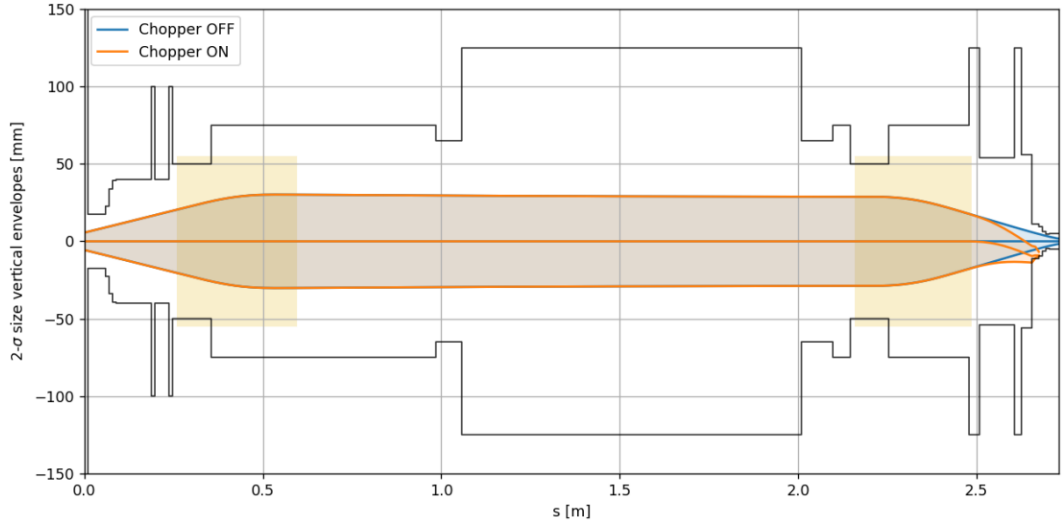

**Fig. 10:** Beam optics of the LEBT showing the focusing of the beam by two solenoids. The chopper allows to deflect the beam on a conical dump before the RFQ.

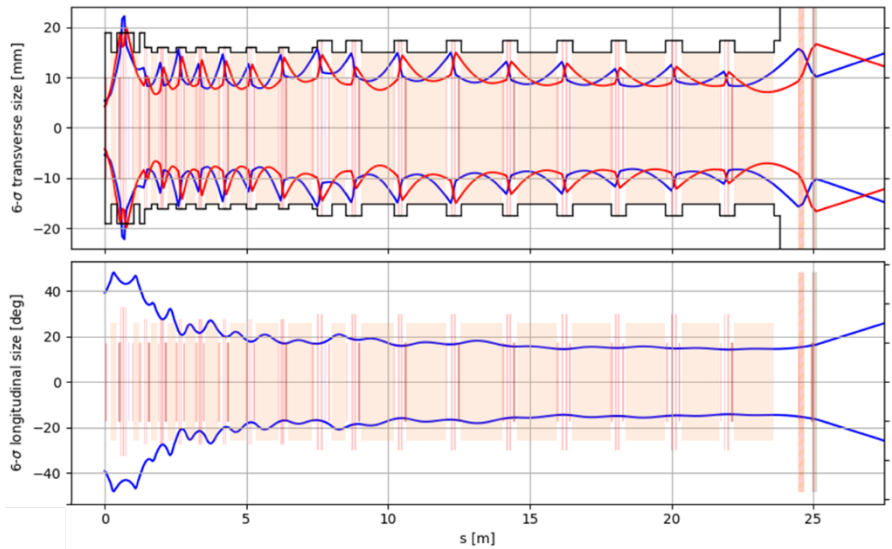

**Fig. 11:** Transverse (top) and longitudinal (bottom) 6σ beam sizes in the CH-section.

Since a failure in one cavity cannot be compensated by adjacent cavities, a complete second injector is required to ensure high reliability. MYRRHA's final design includes one straight injector with a second one connected via a dog-leg. There are two possible configurations for this switching:

- Switching Dipole with a 1s ramp duration
- A Kicker-Septum configuration enables millisecond switching, keeping both injectors fully qualified and operational. Switching is simply a matter of adjusting timing system events. This also allows a failed injector to be gradually reintroduced and retuned.

The merging section can also deflect the beam to a 17 MeV beam dump, facilitating beam tuning.

## 6.2 SC-linac

The linac is composed of cryomodules, each housing 2 to 5 cavities (depending on the energy range), separated by warm sections at room temperature. These warm sections include a quadrupole doublet, orbit correction, a BPM, and vacuum equipment. The 6σ beam optics (both transverse as longitudinal) in the low-beta section of the MYRRHA linac is presented in Fig. 12, together with the phase advances in this section after each lattice period of 3.04 m. The coloured boxes in the background represent the periodic optical elements (quadrupoles and cavities).

Contrary to the CH cavities, the requested RF power to accelerate the beam in the low-beta section of the MYRRHA accelerator is much closer to the actual beam power added to the beam by each cavity, giving the high Q-value of the superconducting cavities. This is shown in the top part of Fig. 15, where the black boxes represent the beam power added by each cavity, the red boxes the requested RF power and the blue lines indicate the available RF power by the solid state amplifiers.

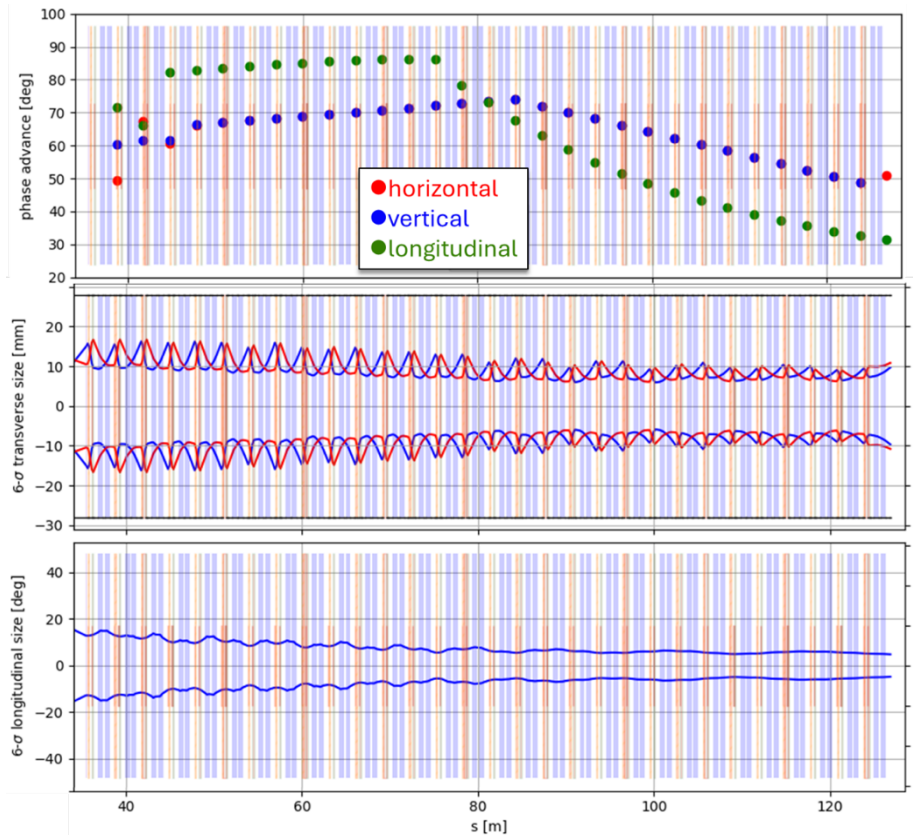

**Fig. 12:** SC-linac beam optics using the low-beta section as example. Top = the transverse and longitudinal phase advances, middle = 6σ transverse beam size and bottom = 6σ longitudinal beam size.

The main longitudinal tuning parameters follow standard beam-dynamics principles for high-power ion linear accelerators [5],[6]:
   – The synchronous phase at the start of the superconducting linac is set to a sufficiently low value (−45°, see Fig. 14) to provide a large longitudinal acceptance. Figure 13 shows the simulated longitudinal acceptance of the MYRRHA SC linac (low-beta section only) in grey with the incoming beam phase space on top with a logarithmic intensity colour plot.
   – The zero-current longitudinal phase advance is maintained below 90° per lattice period to avoid structure- and space-charge-driven resonances [7] (see top of Fig. 12).

- The phase advance per meter is kept as smooth as possible throughout the linac to reduce mismatch risks and ensure a lattice that remains as current-independent as possible (see top of Fig. 12).

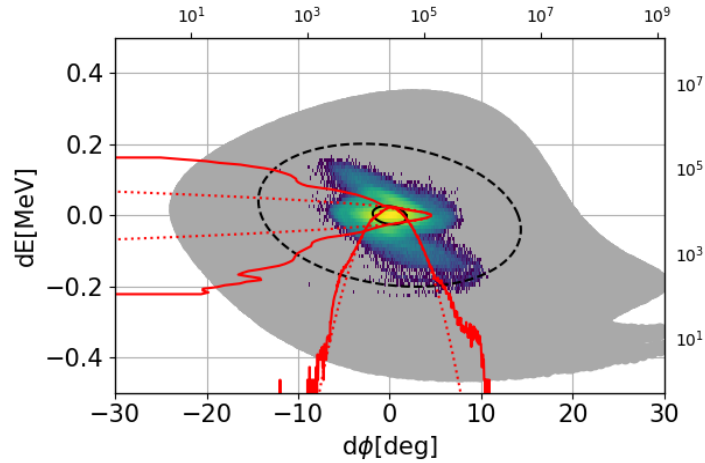

**Fig. 13:** Acceptance plot of the longitudinal phase space. The grey area shows the acceptance where particles are transported and accelerated successfully through the full linac.

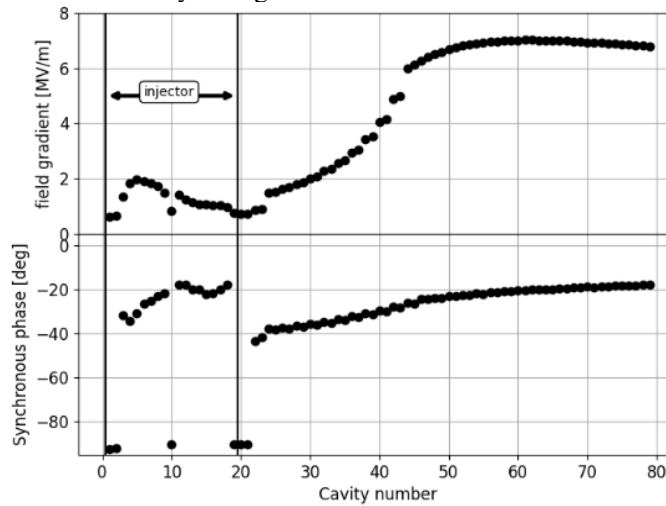

**Fig. 14:** (top) the field gradient in the CH cavities (injector) and the SC cavities along the MYRRHA linac (up to the low-beta section). (bottom) the synchronous phases in the injector and low-beta section.

Since the cavities are identical, it is possible to compensate for the failure of a single cavity by adjusting the fields in its neighboring cavities. While this strategy is already successfully employed in operational accelerators, ADS accelerators allocate a larger RF overhead per cavity and aim to execute such reconfigurations more rapidly. Two main compensation schemes are considered:
- Local compensation, which aims to minimize the number of cavities whose settings are modified.
- Global compensation, which aims to minimize the required RF overhead by distributing the compensation across a larger set of cavities.

An example of a local compensation of a failed cryo-module is shown in the lower part of Fig. 15. In this case, the 10$^{th}$ cryomodule in the low-beta section failed and the lacking energy gain is added by 2 cryo-modules before and after the failed cryo-module.

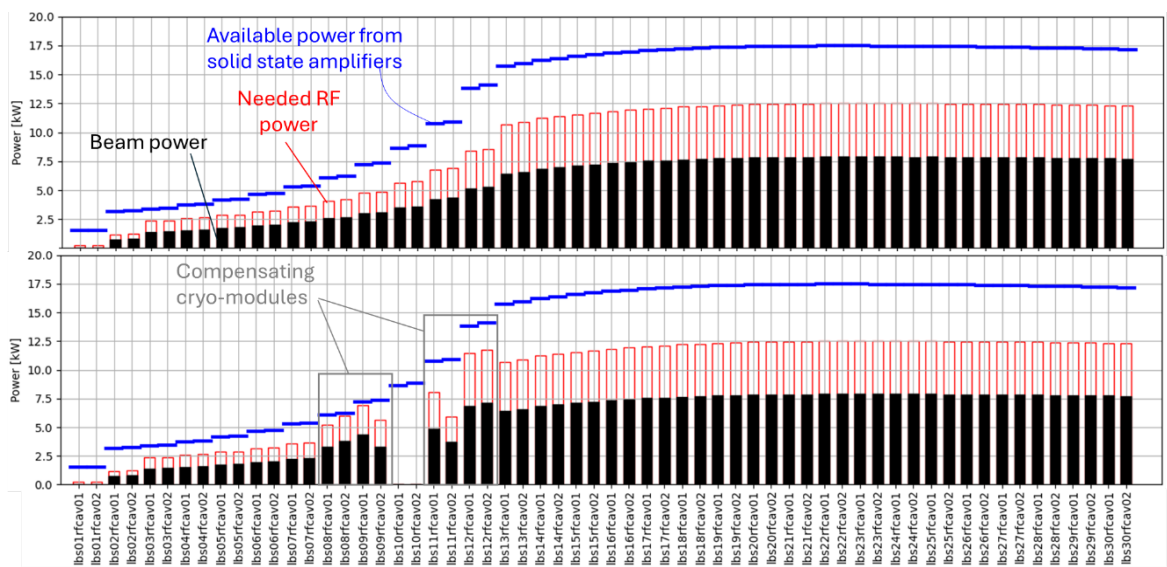

**Fig. 15:** The beam power (black boxes) and the required RF power from the solid state amplifiers (red boxes) in the low-beta section of the MYRRHA SC linac. The blue lines indicate the RF overhead foreseen in the solid state amplifiers to compensate for failures of cavities or cryomodules. The lower plot shows a situation where the 10$^{th}$ cryomodule failed and is compensated by 4 neighbouring cryo-modules.

Any such adjustment must be accompanied by corresponding quadrupole retuning in the affected regions, as well as a refreshed orbit correction.

# 7 Other critical technologies

In addition to the RF cavities, some further key technologies are highlighted in the following.

## 7.1 Cryogenic System Overview

The cryogenic system for the 100MeV stage of MYRRHA manages a total helium inventory of 700 kg. It consists of three main subsystems:

### 7.1.1 Cryo-plant

The cryo-plant provides the cooling power required to liquefy and maintain the helium at operating temperature. Achieving stable and reliable operation at these conditions is a major challenge.

The cooling of the total heat load (3.5 kW at 4.5 K, including: 900 W at 2 K which is ≈70% of the total thermal load) consumes about 1MW of electrical power.

### 7.1.2 Cryo-distribution System

This subsystem transports helium between the cryo-plant and the cryogenic users. A valve box is installed for each cryomodule and taps into the cryogenic backbone. Each valve box contains a heat exchanger and a Joule–Thomson (J–T) valve to expand and cool helium down to 2 K.

### 7.1.3 Cryomodule

The cryomodule provides the environment required for stable SRF cavity operation and ensures mechanical and magnetic stability.

The cryomodule keeps the SRF cavities at 2 K using:

- An insulating vacuum ($10^{-6}$–$10^{-7}$ mbar)
- Multi-layer insulation (MLI) foils
- A thermal shield at intermediate temperature to intercept radiative heat loads.

It maintains precise cavity alignment with respect to the beam axis during cool down and minimizes mechanical vibrations (microphonics) that could detune the cavities

It includes a magnetic shield to suppress external magnetic fields at the cavity surface.

The Cold Tuning System mechanically deforms the cavity to control its resonance frequency using two actuation mechanisms:

- Slow, large-range tuning: via an electromechanical motor
- Fast, small-range tuning: via piezoelectric actuators.

In most cryomodules, the motor is located inside the cryomodule vacuum vessel. For MYRRHA, however, frequent and rapid detuning is required for testing, which motivated a new design: The motor is placed outside the vacuum vessel and the torque is transmitted through a power-train mechanism.

Since niobium's ductility changes significantly during warm-up, the tuning system must disengage during temperature rise to avoid permanently deforming or damaging the cavity.

The CW RF operation implies a relatively high dynamic heat load compared to the static one MYRRHA: 9W/9W (compared to the pulsed ESS Spoke cryomodule: 1W/16W).

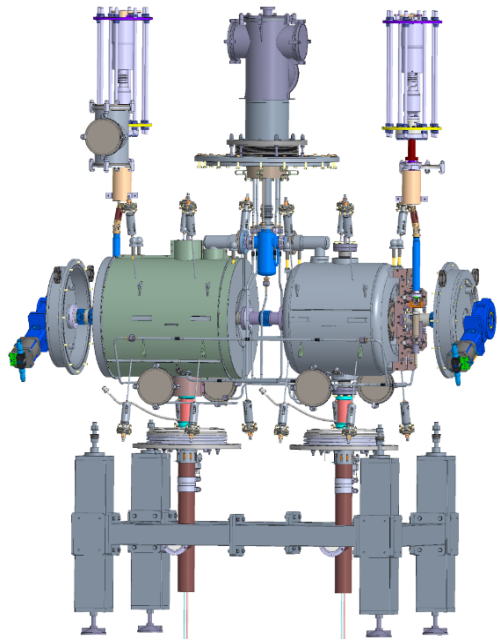

**Fig. 16:** Illustration of the MYRRHA cryo-module with the outer vacuum vessel removed.

## 7.2 Solid state amplifiers

The required RF power - in the lower hundreds of MHz range and at power levels of several tens of kilowatts per cavity - is very well suited for solid-state RF amplifiers (SSAs). Modern SSAs offer several advantages over tube-based systems, most notably:

- High intrinsic reliability:
  Solid-state devices operate far below their maximum ratings and have excellent mean time between failures (MTBF).
- Inherent fault tolerance:
  SSAs are typically built from many ≈1 kW amplifier modules, each containing multiple transistors. If a single transistor or even an entire 1 kW module fails, the system can automatically compensate by redistributing power among the remaining modules. Operation can continue with only a small reduction in output power, greatly improving availability.
- Simplified maintenance:
  Faulty modules can be hot-swapped or replaced without impacting the rest of the amplifier.

The overall wall-plug efficiency of such amplifier systems is approximately 60%, although the exact value depends on operating point, cooling system, and transistor technology.

Because the required RF power can vary significantly during cavity detuning and fault-compensation scenarios, it is essential for the SSA to include a drain-voltage adaptation. This feature allows the amplifier to adjust its DC supply voltage according to the demanded RF output power, which provides improved efficiency at partial load and reduced thermal stress on the transistors.